\documentclass[10pt, conference, compsocconf]{IEEEtran}

\usepackage{color}
\usepackage{listings}
\usepackage{url}
\usepackage{array}
\usepackage{graphicx}
\usepackage{amssymb}
\definecolor{mygray}{rgb}{0.5,0.5,0.5}
\begin{document}

\title{Reproducing Scientific Experiment with Cloud DevOps}

\author{\IEEEauthorblockN{Feng Zhao}
\IEEEauthorblockA{Tsinghua University \\ Beijing, China \\ zhaof17@mails.tsinghua.edu.cn}
\and
\IEEEauthorblockN{Xingzhi Niu}
\IEEEauthorblockA{University of Washington \\ Tacoma, USA \\ nxz18@uw.edu}
\and
\IEEEauthorblockN{Shao-Lun Huang}
\IEEEauthorblockA{Tsinghua University \\ Shenzhen, China \\ shaolun.huang@sz.tsinghua.edu.cn}
\and
\IEEEauthorblockN{Lin Zhang}
\IEEEauthorblockA{Tsinghua University \\ Shenzhen, China \\ linzhang@tsinghua.edu.cn}
}
\maketitle
\begin{abstract}
The reproducibility of scientific experiment is vital for the advancement of disciplines based on previous work. To achieve this goal, many researchers focus on complex methodology and self-invented tools which have difficulty in practical usage. In this article, we introduce the Cloud DevOps infrastructure from software engineering community and shows how it can be used effectively for heterogeneous agents to reproduce experiments for computer science related disciplines. DevOps can be enabled using freely available cloud computing machines for medium-sized experiment and self-hosted computing engines for large-scale computing, thus powering researchers to share their experiment result with others in a more reliable way.
\end{abstract}
\begin{IEEEkeywords}
cloud infrastructure; DevOps; scientific workflow management
\end{IEEEkeywords}

\section{Introduction}
With the development of cloud computing, computational scientific experiments encompass more disciplines and are much more complex than before. Cloud computing offers many convenient architectures which has been proven useful in scientific experiment scenario.  For example, some highly paralleled experiments can be done on Cloud Serveless with affordable cost \cite{niu2019leveraging}. There are other emerging domains like Big Data which are related to computational scientific experiment. These experiments require more dedicated toolchains, specific workflow and expensive computational resources, which put new challenges on experiment reproducibility.

Besides these traditional challenges, many experiments, especially in hardware related domain, are conducted on
embedded devices. Reproducing these experiments are more difficult than simulation-only ones since other researchers may not share exactly the same model \cite{report2017}. The scale, running time and platform of experiments vary so greatly that it is hard to find a
one-fit-all solution for the experiment reproducibility problem.

To solve the reproducibility issue, there are three kinds of approaches: tools-oriented, platform-oriented and methodology-oriented. For tools-oriented approach, there are many useful tools, which can capture the running environment information or storing the experiment results \cite{greff2017sacred}. The environment
capturing functionality is designed since preserving provenance of experiments is essential for reproducibility \cite{freire2018provenance}.
These tools are valuable but may suffer from bad maintainability and difficult configuration.
Indeed, they are made by domain-specific scientists, not by experienced full-time software engineers. These tools can store the experiment results, which requires users to configure the database locally but researchers may not be experienced in database. Besides, the local data is difficult to share. Furthermore, most of these tools are not programming language and computing architecture neutral, which means researchers cannot use them in other environment. Still, something is better than nothing if researchers are able to use these tools to manage their experiment.

For platform solution, traditionally containerization is used. In recent years, it has been shown that cloud computing is suitable for scientific research purpose \cite{Howe12} and cloud-based hardware is emerging \cite{cloudhard}. Configuration on cloud environment from scratch is difficult for inexperienced researchers and it is better to use specific cloud service for research purpose. For example, we have Code Ocean or other commercial cloud systems \cite{perkel2018data}, which can tackle the reproducibility problem to some extent. However, their free tier is mean and researchers probably are not willing to pay for extra computational resource. Budget limitation is an important factor when it comes to buying cloud computing resources.

Methodology in reproducibility usually discusses the general principles \cite{stodden2014best} or combines tools and platforms to explore the best practice \cite{QashaCW16}. Generally speaking, methodology is hard to follow as they tend to be ideal and the researchers may not be familiar with or have the ability to setup the toolchain used.  

All the above three aspects have pros and cons for experiment reproducibility. The key is how to combine the three aspects to make the best use of their advantages. This is what DevOps tries to solve. This idea is not newly proposed. Boettiger gives a try using Docker container for experiment reproducibility \cite{Boettiger15}.
He also mentioned the DevOps philosophy and acknowledged its limitations.
There are also hardware research projects which borrow the ideas of DevOps to conduct sophisticated experiments \cite{chwalisz2019walker}. These previous researches are valuable but they are limited to specific domain and local environment.

Except for local system, there are also dedicated system on the cloud, which tries to solve domain specific problem and is related with experiment reproducibility. \texttt{Devops@mech} is developed for a certain institute, which is based on DevOps methodology \cite{philips2019devops}. For public service, we have RAMP for data science domain \cite{kegl2018ramp} or VCR for computational results indexing purpose \cite{GavishD12}. Though these services are available when corresponding paper is written, they are unavailable now.
Everest, claimed to simplify the use of clouds for scientific computing, is still available but users are required to attach their own resources before actually using it \cite{VOLKOV2017112}.
Just like existing tools, these lab-made services suffer from bad maintenance.

From the above analysis, we see that previous combination of DevOps with scientific experiment has some shortcomings. In this article, we propose Cloud DevOps approach, which uses DevOps from cloud service point of view. It has the following advantages which are not present simultaneously in previous approaches:
\begin{itemize}
	\item high availability of the service and well maintenance of the infrastructure,
	\item easy to use and flexible to configure,
	\item unlimited usage and rich computing resource
\end{itemize}

The contribution of this paper lies mainly at three aspects. Firstly, we argue that Cloud DevOps can be helpful to reproducibility for general scientific computing experiments. Secondly, we show how to achieve this by designing a
DevOps workflow. Finally, we demonstrate how to apply our workflow to practical experiments.

In the following sections, we will give an introduction to Cloud DevOps and show the feasibility to incorporate existing tools in Cloud DevOps. We then investigate the reproducibility problem with some proof-of-concept examples. These examples take the advantage of Cloud DevOps while integrating old toolchains. We believe Cloud DevOps can help researchers be more productive in their experiments and make it easier for others to follow their research.
 
\section{INFRASTRUCTURE}
Originally, DevOps refers to the software engineering approach to automate the process of building and deploying software product, which is summarized by its two core components ``Continuous Integration and Deployment (CICD)'' \cite{leite2019survey}.
DevOps service (server) can be self-hosted or centrally hosted. In either way, it requires some other computing machines (called agents or runners) to actually run the submitted jobs. Usually the jobs are not submitted manually to DevOps server but the submission is triggered by an update to the code repository. 
DevOps server is quite complex and self-hosted solution is not suitable for sharing results with others. Therefore, it is preferred to use public cloud DevOps service, which provides some free time-unlimited cloud computing agent. Self-hosted computing agent can also be used if public provided agent is not suitable to reproduce the experiment due to computing resource limitation or architecture restrictions. In this article, we only consider cloud hosted DevOps service and call them Cloud DevOps for short.

There are some similarities between Cloud DevOps and Everest infrastructure \cite{VOLKOV2017112} . Both of Cloud DevOps and Everest allow dynamic provision of computing resources from public cloud service provider and support computing agents attached by users. The computation can be triggered by user when the button is clicked via web interface.
However, Everest suffers from problems mentioned in the last section. From the workflow management point of view, Cloud DevOps is similar to the Pegasus system \cite{Pegasus}. While the latter is more suitable for large-scale distributed computing management, Cloud DevOps is scalable and covers the need from small experiment to large-scale experiment as well.

There are many freely available Cloud DevOps service providers for open source project which greatly power individual developers and open source community. {\bf Table} \ref{tab1} gives some famous providers with the list of their features.
\begin{table}
\caption{Comparison of Cloud DevOps provider (until 2020)}
\label{table}
\small
\begin{tabular}{|@{\hspace{0.1em}}m{0.9cm}|@{\hspace{0.1em}}>{\centering}m{0.9cm}@{\hspace{0.9em}}|@{	\hspace{-0.1em}}>{\centering}m{0.9cm}|@{\hspace{0.2em}}>{\centering}m{0.8cm}|>{\centering}m{0.8cm}|>{\centering}m{1.0cm}|c|}
\hline
& 
{\scriptsize AppVeyor }& 
 {\scriptsize Azure pipelines} & {\scriptsize CircleCI } &  {\scriptsize GitLab CICD} & {\scriptsize GitHub Action}  & {\scriptsize Travis} \\
\hline
 {\scriptsize Platform} & {\scriptsize Windows Linux} & All & All & Linux docker & All & All\\
\hline
 {\scriptsize Parallel} & 1 & 10 & 4 & 8 &  20 & 5\\
 \hline
 {\scriptsize  Selfhost } & Y & Y & N & Y & Y & N\\
 \hline
 {\scriptsize Artifact} & N & Y & Y & Y & Y & N\\
 \hline
\end{tabular}
\label{tab1}
\end{table}

In Table \ref{tab1}, ``Platform'' row summarizes what kinds of operating system (OS) are supported; ``Parallel'' row represents the maximum number of allowed parallel jobs; ``Selfhost'' row means whether the service provider supports connection of self-hosted agent; ``Artifact'' row specifies whether it supports preservation of job artifacts. Our approach is not limited to a specific DevOps cloud service provider. Instead, we focus on the general methodology, which is applicable to almost any service provider.

We call a DevOps infrastructure cross-platform (\texttt{Platform = All} in Table  \ref{tab1}) if it supports Windows, MacOS and Linux 
OS.
Cross-platform is an important topic in software engineering. For scientific community, most research experiments can only be reproduced on specific version of one operating system. This is OK since researchers may not have machines of other operating systems, or they have no time to make their code run on different platforms. A recent study found a flaw of Python script in an article published on Nature which produces different results on different OS \cite{bhandari2019characterization}. This incident can be avoided if researchers test their experiment code on different OS. Cloud DevOps provides easy configuration for different environments and researchers are encouraged to test their code on different OS without learning too much new knowledge and spending too much time. To the least extent, researchers can choose the most similar environment on the cloud to their local development environment and make the experiment runnable on cloud. To the largest extent, it is beneficial if newly developed algorithms and experiments can be run on more platforms.

Parallelism is a valuable capability of Cloud DevOps. In software engineering community, it is often used to run different tests in parallel.
Artifacts are build product which are ready to be deployed to other places.
Some DevOps service provider give the opportunity to save partial artifacts permanently. For scientific experiment scenario, independent experiments can be run in parallel jobs and the results (like figures) can be saved automatically for each job and viewed by public.

Similar to cross-platform notation, we call a platform ``cross-architecture'' if this platform supports more architectures besides X86. Currently, some Arm architecture variants are experimentally supported for a few platforms. For example, Travis gives the user some freedom to build the software
for ARM64 while GitHub Actions support self-hosted ARM device \cite{travis-arm, github-selfhosted}. In the future, we believe there will be more progress in the ``cross-architecture'' domain for Cloud DevOps. 

Cloud DevOps uses configuration file to determine the running environment and workflow instructions. 
Usually the configuration file is written in \texttt{YAML} format. Different Cloud DevOps providers have different schema in this format, but they all do the same thing. Below we give a short introduction of how to configure Cloud DevOps to run the experiment.
\subsection{Choosing Environment for Agent}
Users first choose the actual running environment of their code. Usually, it is the combination of the following items:
\begin{enumerate}
\item virtual machine or docker container,
\item public cloud service or local runner,
\item programming language and version.
\end{enumerate}

For example, on Travis users can have  \texttt{Ubuntu 16.04 Python 3.6} environment by simple requires it in the following way:
\begin{lstlisting}[caption={environment configuration}]
os: linux
dist: xenial
language: python
python: 3.6
\end{lstlisting}

In this configuration, we use the Linux virtual machine provided by cloud service. We also fix the version of certain software. 
Such shortcut makes installing dependency in later workflow management much easier as we do not need to install \texttt{Python} or other software programs manually.

Besides virtual machine, many DevOps infrastructures support Docker containers as well, which provides more flexible way to configure the environment. Generally speaking, virtualization is better than host OS for experiment reproducibility \cite{Howe12}. Hence, Cloud DevOps can do a good job by providing out-of-the-box virtual machine.

\begin{figure}[!ht]
\includegraphics[width=18.5pc]{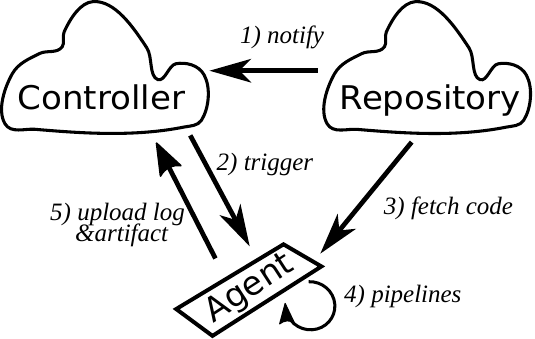}
\caption{Interaction of DevOps server with agent and code repository}\label{fig:principal}
\end{figure}

Usually Cloud DevOps is used in cooperation with a source code repository. The system diagram in Fig. \ref{fig:principal} shows the interaction among the computing agent, the Cloud DevOps server and the code repository. When the source code is updated, Cloud DevOps server triggers the computing agent to run the build. The agent fetches the latest code automatically and uploads the log file and artifacts to the server after it finishes tasks. The artifacts can be figures of experiment result which will be included in the paper.
The code repository is used for version control and each running of the agent produces a log file. The source code version and the log file are 1-to-1 correspondent. Inspecting the public available log and its corresponding source code helps others to reproduce the same result using public available build machine or
self-hosted agent.

Though we can use the cloud computing resources for unlimited time, the parallel ability on algorithm level is restricted and special computing device (like GPU) or programming model (MPI) is missing. Thus, it is necessary to use self-hosted environment to run complex scientific experiment. Fortunately, many Cloud DevOps service provides the local agent option. By installing a client software on one's own machine, it is possible to empower the advantages of Cloud DevOps without losing the computing ability of self-hosted devices.
\subsection{Describe Workflow for Agent}
\begin{figure}[!ht]
\includegraphics[width=18.5pc]{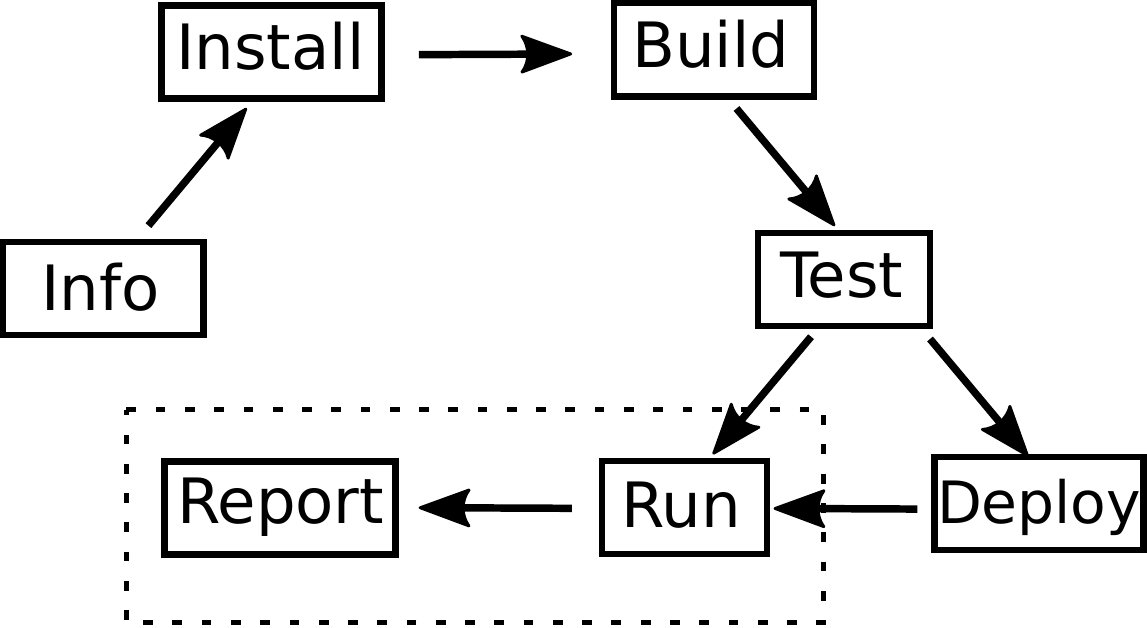}
\caption{CICD pipeline illustration. The steps within dashed boxes are specific stages for scientific experiments. }\label{fig:cicdworkflow}
\end{figure}

After choosing the running environment, users should determine how to execute their code sequentially. The basic workflow can be summarized in Fig. \ref{fig:cicdworkflow}.

The first few steps are common. We need to capture enough information of the running machine (\texttt{Info}) and install necessary software dependencies (\texttt{Install}). Then we build our source code to binary executable (\texttt{Build}) and run some test to verify whether it works for simple cases (\texttt{Test}). In software engineering community, DevOps ends with the deployment step. But for scientific experiment, the story just begins after packing your algorithm to a reusable package. Therefore, we use dashed boxes to emphasize the unique steps for scientific experiment in Cloud DevOps infrastructure. After the test, we begin to run the experiment (\texttt{Run}) and finally the result needs to be collected and further processed to produce the artifacts (\texttt{Report}).

For \texttt{Info} step, it is automatically done by DevOps server. For other steps, shell scripts are used to tell the running machine how to install, build and run the code. Not all steps are necessary. For example, no \texttt{Build} step is needed for interpreted programming language. Suppose a researcher writes his experiment code using Python programming language, then he can write his workflow as follows:
\begin{lstlisting}[caption={workflow description}, label={lst:wd}, basicstyle={\small}]
install: 
  - pip install -r requirements.txt
script: # run experiment
  - python main.py
\end{lstlisting}

In the above workflow description, \texttt{Build}, \texttt{Test} and \texttt{Deploy} steps are omitted. This is common for many researchers: usually they do not test and deploy their code. That's OK as long as the experiment results are all right. Still, it is better to do some test and deployment task. Deployment makes other researchers easier to compare their results with your method without copying your code to their own repository and modify your code to fit their need.

Since the configuration of Cloud DevOps is transparent to all users and the mechanism of it is totally determined by configuration file and a specific version of source code. Other researchers can trust the output logs and artifacts of DevOps as evidence of experiment reproducibility. Rerunning the code is quite easy: just use the same service provider and the code can be run under a different account after replicating the code repository. We acknowledge that this convenience is not applicable to self-hosted agent. For self-hosted agent usually the environment configuration part is not written in a file but determined by the host OS of the attached agent. This made reproducibility less easy. Still, the logs and artifacts are available to be examined by public since they are uploaded to public Cloud DevOps server from local agent. To make the story of experiment reproducibility complete,
we encourage the researchers to run a partial and small scale experiment on public agent and run their full experiment on self-hosted agent using the same code.

\section{CASE STUDIES}
In the previous section, we briefly overview the common practice in DevOps and how it can be related with scientific experiment reproducibility. Different domains may still face different problems in practice. In this section, we use experiments from the domain of graph computing, bioinformatics and scientific software development to show how Cloud DevOps can be used to solve the reproducibility problems. All experiments follow our proposals and can be checked from \url{https://github.com/Leidenschaft/experiment_reproducibility_collection}. We believe Cloud DevOps can be used in experiments of other domains as well. 

\subsection{Using public agent}
Generally, if researchers develop a new algorithm for a specific domain, the workflow shown in Fig. \ref{fig:cicdworkflow} can be further decomposed into two phases:
algorithm library build phase and experiment running phase. The output of the first phase is the reusable library which is one of input to the second phase. Using DevOps in the first phase is nearly identical to how DevOps is used in software community.  The code can be tested against different environment and 
the reusable library can be deployed to public available package repository.
 
Following this two-phase philosophy, we consider a simple triangle-counting algorithm and apply it to considerably large graph. 
In the first phase, we compile the code and deploy the package to Ubuntu PPA. We also demonstrate the code can be compiled and run successfully on Windows by using AppVeyor. In the second phase we just install the deployed package and run the actual experiment on the agent provided by Travis, which has 2 CPU cores and 7.5 GB memory. The log of this experiment can be checked publicly on Travis, which shows
our program consumes 3.1 GB memory in peak and finishes in 4.3 minutes. 

\subsection{Using self-hosted agent}
\begin{figure}[!ht]
\centerline{\includegraphics[width=18.5pc]{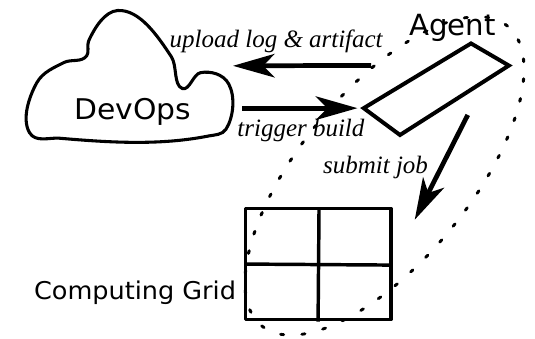}}
\caption{Using self-hosted HPC to connect DevOps server. The dashed ellipse part is self-hosted resources.}\label{fig:selfhosted}
\end{figure}

Cloud DevOps public agent provides convenient environment for general purpose task but is not suitable to run long-time experiment due to time limitation for a single run. For this kind of experiment, self-hosted agent should be used. Self-hosted agent means any computing devices provided by the user, which can be laptops, workstations, lab bare-metal server, paid cloud virtual machines etc.
For our triangle-counting experiment, we use our lab bare-metal server to run the larger experiment. Since we use OpenMP to do some parallelism on algorithm level, the multi-core CPUs on the server help a lot in accelerating the experiment. We also choose a larger dataset which requires 18 GB peak memory to consumes it. We use GitLab as the DevOps service provider. Our self-hosted agent is a head node in an HPC cluster, and we use it to submit the job to computing nodes.
The relationship is illustrated by Fig. \ref{fig:selfhosted}. Using agent to submit job has extra advantages that the running logs are preserved in a continuous way without messing things up. Since dependency of experiment can be compiled and prepared beforehand, generally only dashed box workflow in Fig. \ref{fig:cicdworkflow}
is executed in self-hosted agent. Going through the whole pipeline of DevOps costs extra time but makes the experiment more reliable.

Each computing node of ours has 56 CPU cores, 256 GB memory, and we need around 5 hours to run this experiment. The log of each running for self-hosted agent is available publicly on GitLab.

\subsection{Incorporating other cloud infrastructure}
Cloud DevOps is not exclusive and can incorporate other infrastructure as well.
To demonstrate this, we use
an experiment which applied sequence alignment algorithm to protein sequences.
Originally the experiment is run on AWS Lambda, which is a Serverless infrastructure provided by Amazon \cite{niu2019leveraging}.
Serverless infrastructure allows many concurrent experiments to run in isolated environment and
we need a client to coordinate them. The client program we used is run on public agent provided by Microsoft Azure. We first compile the experiment source code and deploy it to AWS Lambda platform using Travis. Then the client program is run to invoke Lambda functions and collect the experiment results.
The overall experiment finishes in 2 minutes and produces public viewable logs on Azure pipelines. 

From this example we also notice that Cloud DevOps is not bound to a specific provider. We can use DevOps provided by Microsoft to trigger the experiments deployed on Amazon.

\subsection{Running experiment on other architecture}
In this subsection, we demonstrate Cloud DevOps is able to track the experiment for
self-hosted armhf architecture. We have used a Raspberry Pi  3B Model with Raspbian OS connected
with GitHub Actions. The task is to compile a newer version of slurm, which is a cluster management
toolkit. It takes about 2 hours to finish one run, which is slower than the same task on X86 architecture. The console output of each build on our Rasberry Pi is captured and uploaded to GitHub. Though hardware environment may not be exactly same, other researchers can still benefit from viewing the logs to reproduce or customize their own experiment on other architectures.

\section{CONCLUSION AND FUTURE WORK}
DevOps infrastructure is actively maintained by software engineering community and evolves towards better usability. It will be beneficial for scientists to incorporate DevOps into their daily research. Researchers can run experiment demo, medium-sized or partial experiments directly on public DevOps service and complete their computing intensive or hardware-related experiments using self-hosted agent. Besides running experiment code, the full research cycle is complex and may require retrieve different data from distributed repositories, pre-processing and post-processing steps. Such process has iterative nature, which is similar to DevOps philosophy. Therefore, it is meaningful to investigate how to incorporate research cycle with DevOps cycle to improve the research quality. Currently, it is unknown that beyond scientific software development, how much the accessibility of DevOps by scientific community is. Since DevOps is easily configurable and compatible with existing tools, we believe it will sweep more disciplines in the future.

\section{ACKNOWLEDGMENT}

This work is supported in part by the Natural Science Foundation of China under Grant 61807021, in part by the Shenzhen Science and Technology Program under Grant KQTD20170810150821146, and in part by the Innovation and Entrepreneurship Project for Overseas High-Level Talents of Shenzhen under Grant KQJSCX20180327144037831.

\bibliographystyle{IEEEtran}
\bibliography{exportlist}

\end{document}